%% file: EchoDebugging_IWST_HalUpload.tex
\documentclass[sigplan, 10pt, authorversion]{acmart}

\usepackage{multirow}
\usepackage{pifont}

\usepackage[T1]{fontenc} 
\usepackage[utf8]{inputenc} 
\usepackage{times}
\usepackage[scaled=0.85]{helvet}
\usepackage{graphicx}
\usepackage{ifthen}
\usepackage{xspace}
\usepackage{alltt}
\usepackage{latexsym}
\usepackage{url}            
\usepackage{amsfonts}
\usepackage{amsmath}
\usepackage{stmaryrd}
\usepackage{enumerate}
\usepackage{xspace}

\input{macros}
\AtBeginDocument{%
	\providecommand\BibTeX{{%
			\normalfont B\kern-0.5em{\scshape i\kern-0.25em b}\kern-0.8em\TeX}}}

\setcopyright{acmcopyright}
\copyrightyear{2020}
\acmYear{2020}
\acmDOI{10.1145/1122445.1122456}

\acmConference[IWST20]{IWST20: International Workshop on Smalltalk Technologies}{September 29th and 30th, 2020}{Novi Sad, Serbia}
\acmBooktitle{IWST20: International Workshop on Smalltalk Technologies,
	September 29th and 30th, 2020, Novi Sad, Serbia}
\acmPrice{15.00}
\acmISBN{978-1-4503-XXXX-X/18/06}

\begin{document}

\title
[First Infrastructure and Experimentation in Echo-debugging]
{First Infrastructure and Experimentation in Echo-debugging}

\author{Thomas Dupriez}
\email{tdupriez@ens-paris-saclay.fr}
\affiliation{
	\institution{\textit{Univ. Lille, CNRS, Centrale Lille, Inria}\\
		\textit{UMR 9189 -- CRIStAL}}
	\city{Lille}
	\country{France}}

\author{Steven Costiou}
\email{steven.costiou@inria.fr}
\affiliation{%
	\institution{\textit{Inria, Univ. Lille, CNRS, Centrale Lille}\\
    \textit{UMR 9189 -- CRIStAL}}
	\city{Lille}
	\country{France}}

\author{Stéphane Ducasse}
\email{stephane.ducasse@inria.fr}
\affiliation{%
	\institution{\textit{Inria, Univ. Lille, CNRS, Centrale Lille}\\
    \textit{UMR 9189 -- CRIStAL}}
	\city{Lille}
	\country{France}}

\begin{abstract}
As applications get developed, bugs inevitably get introduced.
Often, it is unclear why a given code change introduced a given bug.
To find this causal relation and more effectively debug, developers can leverage the existence of a previous version of the code, without the bug.
But traditional debugging tools are not designed for this type of work, making this operation tedious.
In this article, we propose as exploratory work the echo-debugger, a tool to debug two different executions in parallel, and the  \emph{Convergence Divergence Mapping} (CDM) algorithm to locate all the control-flow divergences and convergences of these executions. In this exploratory work, we present the architecture of the tool and a scenario to solve a non trivial bug.
\end{abstract}

\maketitle
\td{todo: create a githum repo named "echo-debugger" (instead of "DebuggerCommunication") for the echo debugger, and give a link to that in this paper.}
\sd{I would not worry about that yet, it can be done after the paper is accepted.}

\section{Introduction}
\label{sec:intro}

Nowadays, debugging is still a challenge~\cite{Somm01, Zell05a} and sources of hard bugs are numerous~\cite{Pers17a}. 
In addition, the distance between a source code change and the emergence (identification) of a bug can be large, which makes it difficult to understand why a given code change caused a given bug~\cite{Zell99a}. 
However, in some instances, developers have access to an interesting source of information to help them: a previous version of the software not exhibiting the bug~\cite{Wong16a}.

However, having a reference, working, version of the program is not a panacea. Without dedicated support, developers have to run the two versions in separate debuggers, manually step them in parallel, and visually compare the executions.

%


Techniques that compare two similar executions to produce various results already exist~\cite{Wong16a}.
In general, these techniques try to isolate the code fragments that are (suspected to be) responsible of an error.
\textit{Delta debugging}~\cite{Zell99a, Zell02b} takes two versions of a program and finds the smallest subset of code change that turned a given test from green to red.
\textit{Algorithmic debugging}~\cite{Shap82a, Silv11a} tries to isolate faulty code based on how developers assert the outputs of faulty and successful executions.

However, these approaches show limits in two scenarios.
First, we might know exactly which code change introduced the bug and still we cannot understand how it did so.
Second, when we migrate an application from a version of a library/framework to another, the code changes can be gigantic.
Detecting code differences between a working execution (using the old version) and a failing execution (using the new version) might not be useful.
The meaning itself of the code might have changed, things might have been added and others removed.
For instance, when migrating frameworks from a version of Pharo~\cite{Blac09a} to another, the base classes and tools of the language regularly evolve.

In this paper we present \textit{Echo-debugging}: a technique to compare the executions of the failing and working version of the program and find the control-flow differences to help developers debug the program.
%
The contributions of this paper are:
\begin{itemize}
	\item The echo-debugger and its architecture: an interactive debugger to debug two similar executions running in different runtimes.
	\item \emph{Convergence Divergence Mapping} (CDM), an algorithm that fully runs both executions and compares the AST nodes they are executing to build a map of when they diverge and converge in terms of control-flow.  The echo-debugger can then jump the executions to any event of this map the developer wants to inspect.
\end{itemize}

In this paper, we first state our problem of comparing two similar executions, and list the main challenges it involves (Section~\ref{sec:problem}). 
We then expose our solution: the echo-debugger, its architecture, and the CDM algorithm (Section~\ref{sec:contribution} and~\ref{sec:cdm-algorithm}). 
We show a concrete example of how to use the echo-debugger to debug a bug in the Pillar editorial chain code (Section~\ref{sec:example-pillar-configuration-bug}).
We finally discuss our solution (Section~\ref{sec:discussion}), similar works (Section~\ref{sec:related}), future works (Section~\ref{sec:future}) and conclude (Section~\ref{sec:conclusion}).

\section{Comparing Two Similar Executions}
\label{sec:problem}


\paragraph{Problem statement.}
We have as inputs:
\begin{itemize}
	\item Two versions of a program. For example before and after a given commit.
	\item A statement to execute. The developer is interested in how the execution of this statement differs between the two program versions. This will typically be a test that passes in one program version and fails in the other, but it can be any statement.
\end{itemize}


From these inputs, we want a tool that allows the developer to debug both executions of the statement in a comparative fashion, and to understand the impact of the source code differences between the program version.


\paragraph{Difficulties.}
Here, we list the main challenges our solution has to overcome.
\begin{itemize}
	\item \textbf{Challenge 1: Running two versions of the same program in parallel, and controlling them.} Our solution requires the two version of the same program to run in parallel. This is typically not possible in the same runtime. Additionally, our solution needs to control and coordinate the two executions.
	\item \textbf{Challenge 2: Comparing objects across executions.} Although the executions are similar, and they create and manipulate similar objects, the default identity operator ($==$) is entirely unusable because the same objects from different executions are never going to be the same identity-wise.
	\item \textbf{Challenge 3: Comparing control-flows.} The intuitive idea is to find when the executions are \emph{doing different things} and when they are \emph{doing the same thing}. Our solution needs to define these expressions and use these definitions to compare the control-flows of the two executions.
		\begin{itemize}
			\item \textbf{Challenge 3.1: Finding control-flow divergences.} Since the executions both start on the same statement (the one provided by the developer), their control-flows are the same. Our solution needs to step them until their control-flows diverge.
			\item \textbf{Challenge 3.2: Finding control-flow convergences.}
			Challenge 3.1 lets us find the first control-flow divergence, but that may not be good enough to understand the bug. Maybe the control-flow of the executions reconverge on a part of the program, and diverge again later. Our solution needs to recognise if the control-flow of the executions reconverge after a divergence.
		\end{itemize}
	 Combining challenge 3.1 and 3.2 means building a map of when the control-flows of the two executions diverge, converge, diverge again, converge again...

\end{itemize}

\section{The Echo-Debugger}
\label{sec:contribution}


In this section, we describe our solution to debug two similar executions side-by-side: the \textit{echo-debugger} and its architecture.

For clarity in this section, we assume that the developer is debugging a test, which passes in a given version of the program, but fails in another. In general the echo-debugger works to comparatively debug any statement.


\subsection{Echo-Debugging Architecture}
\label{sec:echo-debugging-architecture}

\begin{figure}[!ht]
	\includegraphics[width=\linewidth]{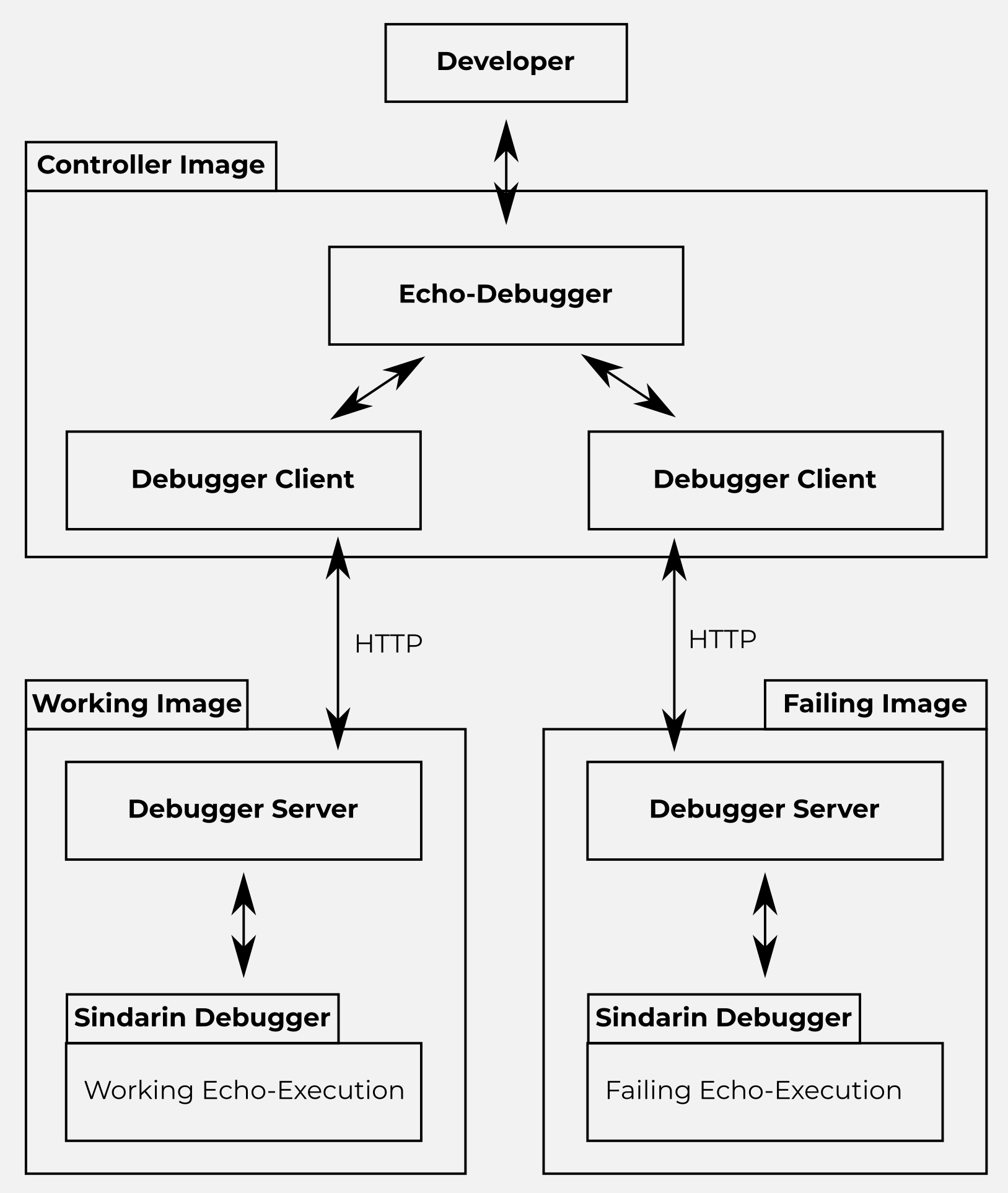}
	\caption{Echo-Debugging Architecture: One controller runtime (image) controls the execution of a failing and working one.  \sd{For later: describing the architecture does not require to be Pharo-centric}}
	\label{fig:echo-debugging-architecture}
\end{figure}

Figure~\ref{fig:echo-debugging-architecture} shows the overall architecture of an echo-debugging setup.

\paragraph{Three Different Runtimes.} 
Because a runtime cannot contain and execute multiple versions of the same code at the same time (challenge 1), the Echo-Debugging architecture is made of three runtimes each one running separately. Each of such runtime runs a different configuration:

\begin{itemize}
	\item \textbf{Working runtime.} This runtime contains the version of the code that works as expected by the developer. For concision, we will call it the \textit{W runtime}.
	\item \textbf{Failing runtime.} This runtime contains the version of the code that \textit{does not} work as expected by the developer. For concision, we will call it the \textit{F runtime}.
	\item \textbf{Controller runtime.} This runtime connects to the other two runtimes to control the executions and collect data. The developer interacts primarily with this runtime during the echo-debugging session. For concision, we will call it the \textit{C runtime}.
\end{itemize}

We refer to the  working and failing runtime as \textit{echo-runtimes}, because they are like echoes of each others: similar, but not exactly the same.

\paragraph{Sindarin Debugger.}
In each echo-runtime, we use a Sindarin debugger~\cite{Dupr19a} to control the execution of the test. Sindarin is a scriptable, UI-less debugger for Pharo. It can be instantiated on an execution, and its API used to inspect and manipulate the execution.

\paragraph{Debugger Client/Server.}
For the communications between the echo-runtimes and the controller runtime, the echo-debugger has a companion package with an HTTP-based client/server communication layer. This layer transmits the Sindarin commands coming from the echo-debugger to the Sindarin debuggers in the echo-runtimes, and transmits back the answers.
Some objects returned by the Sindarin API cannot be serialized/materialized, such as Contexts and Exceptions, because they reference objects that cannot be serialized. We built custom serializations for them, where we instead serialize a dictionary containing the relevant fields of these objects, excluding the unserializable ones.

\paragraph{Echo-Debugger.}
The echo-debugger is what the developer interacts with. It communicates with the Sindarin debuggers in the echo-runtimes via the client/server communication layer. For a more detailed description of the echo-debugger, see Section~\ref{sec:echo-debugger}.

\paragraph{Setup process for an echo-debugging session}
\label{sec:setup-echo-debugging-session}
Finally, here is the list of steps required to setup an echo-debugging session.

\begin{enumerate}
	\item Create three runtimes: \textit{Working}, \textit{Failing} and \textit{Controller}.
	\item Load the working version of the code in the working runtime.
	\item Load the failing version of the code in the failing runtime.
	\item Load the echo-debugger and its communication package\footnote{\url{https://github.com/dupriezt/DebuggerCommunication}} in all three runtimes.
	\item In both echo-runtimes, instantiate a Sindarin debugger~\cite{Dupr19a}  on the execution of the test.
	\item In both echo-runtimes, run a debugger server for the Sindarin debugger.
	\item In the controller image, run a debugger client, connect it to both debugger servers over HTTP, and open its UI.
\end{enumerate}

\subsection{The Echo-Debugger}
\label{sec:echo-debugger}

The echo-debugger is responsible for controlling and analyzing both echo-executions. Once the echo-debugging session is setup, the developer only interacts with the echo-debugger, and not directly with the echo-runtimes.

\begin{figure*}[!ht]
	\includegraphics[width=0.95\linewidth]{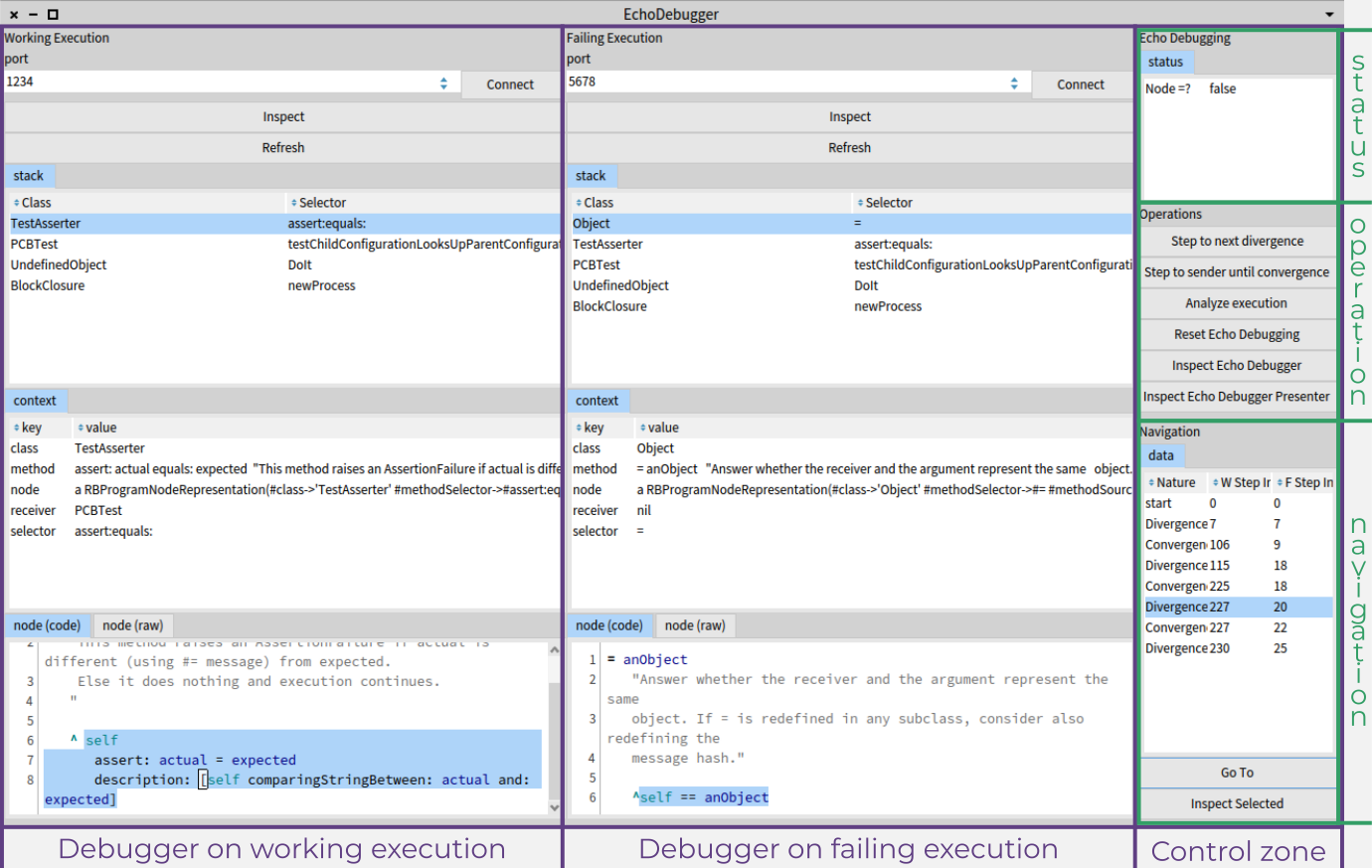}
	\caption{UI of the echo-debugger, after setting up and connecting to the echo-runtimes. The UI is separated into three columns showing, from left to right: the \emph{working} execution, the \emph{failing} execution, and the \emph{control zone}. The control zone is itself separated into three areas: (from top to bottom) the \emph{status area}, the \emph{operations area}, and the \emph{navigation map}.}
\label{fig:echo-debugger-ui}
\end{figure*}

Figure~\ref{fig:echo-debugger-ui} shows the UI of the echo-debugger. It contains three main zones (from left to right):
\begin{itemize}
	\item A debugger on the \textit{working} execution of the test.
	\item A debugger on the \textit{failing} execution of the test.
	\item The \textit{control zone} containing information and commands specific to the echo-debugger.
\end{itemize}

The control zone is separated in three areas:
\begin{itemize}
	\item The \textit{status area} takes the current AST node of the two contexts selected in the debuggers and shows whether they are equal or not.
	\item The \textit{operations area} lists the echo-debugging operations the developer can perform.
	\item The \textit{navigation map} lists the convergence and divergence events between the echo-executions, and allows the developer to step both debuggers to when these events happened in the echo-executions.
\end{itemize}

\paragraph{Remote debuggers.} The echo-debugger features a remote debugger for each echo-execution. These debuggers display information on the echo-executions, such as the call stack and the current piece of code being executed. The developer can use these debuggers to debug the echo-executions as he would normally debug in a standard debugger, with the added benefit of having both executions side-by-side in the same image.

\paragraph{Echo-debugging operations.} The echo-debugger provides operations to control both echo-executions at the same time and step them to potential places of interest: 

\begin{itemize}
	\item \textit{Step both.} Step both echo-executions once.
	\item \textit{Step to next divergence.} To be used when the echo-executions are currently convergent. Step both echo-executions until their next divergence. See Section~\ref{sec:cdm-algorithm} about the CDM algorithm for more details.
	\item \textit{Step to next convergence.} To be used when the echo-executions are currently divergent. Step both echo-executions until their next convergence. See Section~\ref{sec:cdm-algorithm} about the CDM algorithm for more details.
	\item \textit{Analyze executions.} Applies the CDM algorithm described in Section~\ref{sec:cdm-algorithm} to populate the navigation map with all the convergence and divergence events between the echo-executions.
	\item \textit{Restart.} Restarts both echo-executions, to start over.
	\item \textit{Go to.} This operation requires that the navigation map has been populated by analyzing the echo-executions with the CDM algorithm (Section~\ref{sec:cdm-algorithm}). This operation restarts both echo-executions and steps them until they reach the convergence/divergence event that is currently selected in the navigation map. This operation assumes the execution is deterministic
\end{itemize}
The \textit{Step to next divergence} and \textit{Step to next convergence} operations directly address challenges 3.1 and 3.2. \textit{Analyze executions} is a convenience method to automatically repeat these two steps on the entire execution. \textit{Go to} lets the developer inspect each divergence/convergence event. \textit{Restart} and \textit{Step both} give manual control of the parallel executions to the developer for closer inspection.

\section{The CDM algorithm}
\label{sec:cdm-algorithm}

In this section, we explain the CDM algorithm, used by the echo-debugger to spot all the control-flow \textit{divergences} and \textit{convergence} between the echo-executions. We first define what we mean by \textit{convergence} and \textit{divergence}. We then explain the CDM algorithm. We finally detail a special case of the algorithm when looking for a \textit{convergence}, and how we perform the comparison of AST nodes from different runtimes.

\begin{figure}[!ht]
\includegraphics[width=0.4\linewidth]{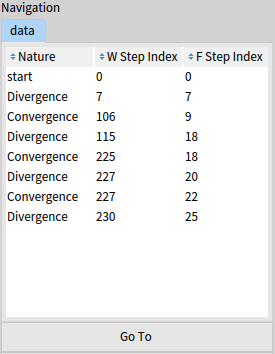}
\caption{Result of the CDM algorithm on the Pillar configuration bug. This is the list of the convergence/divergence events observed during the echo-execution. The left column indicates the nature of the event (convergence or divergence). The middle column indicates the number of steps it has taken the \textit{working} echo-execution to reach this event. The right column indicates the number of steps it has taken the \textit{failing} echo-execution to reach this event.} \label{fig:echodebuggeronpcbbug2}
\end{figure}

The goal of the Convergence Divergence Mapping algorithm (CDM) is to fully run both echo-executions, and build a map of when they diverge and converge in terms of control flow. This map is a list of divergence and convergence events in the order in which they occurred during the comparative execution. Each event stores the number of steps both executions took to reach it. An example of such map is shown in Figure~\ref{fig:echodebuggeronpcbbug2}. Using this map, the echo-debugger is able to re-run the echo-executions up until any divergence/convergence event the developer wants to inspect.

\paragraph{Convergence and divergence of echo-executions.}
We define what we mean by \emph{convergence} and \emph{divergence} as follows.
The idea is that we have two \emph{similar} executions, and we want to know when they are doing the same thing (such as executing the same methods), and when they are not.
At the start, neither echo-execution has executed anything, and they are both about to execute the same statement, provided by the developer. At this stage, they are definitely doing the same thing. We say they are \emph{convergent} at that point.
Then, as the echo-executions progress, at some point, they'll stop doing the same thing. We detect this by comparing the AST nodes they are executing. When they start executing different AST nodes, we say they are now \emph{divergent}.
But we know that prior to that point, they were doing the same thing, so if we let them fully step the current method call, the echo-executions will go back to the caller of that method call, and if at that point they are about to execute the same AST node, we say they have converged. Indeed they were doing the same thing, then they entered a method call in which they started doing different thing, but now that method call is over and they are back to the part where they were doing the same thing.
Now that they have converged, we let them progress until they diverge again, and converge again, etc, until either execution is over.
This definition of \emph{convergence} and \emph{divergence} is the general idea of the CDM algorithm.

\paragraph{The CDM algorithm.}
\td{todo: diagram showing the CDM algorithm in action, if time.}
\sd{there is no time (for later)}
Here is how the CDM algorithm builds a map with the divergences and convergences between the echo-executions. It is mostly a direct translation of the definition of \emph{convergence} and \emph{divergence} we gave in the paragraph above, with the exception of step \textit{2.c.i}. 

\begin{enumerate}
	\item The echo-executions start convergent because they have done nothing yet and are about to execute the same statement, provided by the developer
	\item Repeat until either execution is over:
	\begin{enumerate}
		\item Step to next divergence
		\begin{itemize}
			\item Repeat until the AST nodes the echo-executions are about to execute are \textit{different}:
			\begin{enumerate}
				\item Step each echo-execution once
				\item Compare the AST nodes the echo-executions are about to execute
			\end{enumerate}
		\end{itemize}
		\item Register a \emph{divergence} event in the map, with the number of steps each echo-execution took to reach that point
		\item Step to next convergence
		\begin{itemize}
			\item Repeat until the AST nodes the echo-executions are about to execute are \textit{the same}: \sd{what if there is an infinite loop? (in discussion?)}
			\begin{enumerate}
				\item If the call-stack of both echo-executions do not have the same size, step the echo-execution with the longer call stack until its call stack has the same size as the call stack of the other echo-execution
				\item Otherwise, if the call stack of the echo-executions have the same size, step each echo-execution separately until the size of their call stack is 1 less
				\item Compare the AST nodes the echo-executions are about to execute
			\end{enumerate}
		\end{itemize}
		\item Register a \emph{convergence} event in the map, with the number of steps each echo-execution took to reach that point
	\end{enumerate}
\end{enumerate}

\begin{figure}[!ht]
\includegraphics[width=\linewidth]{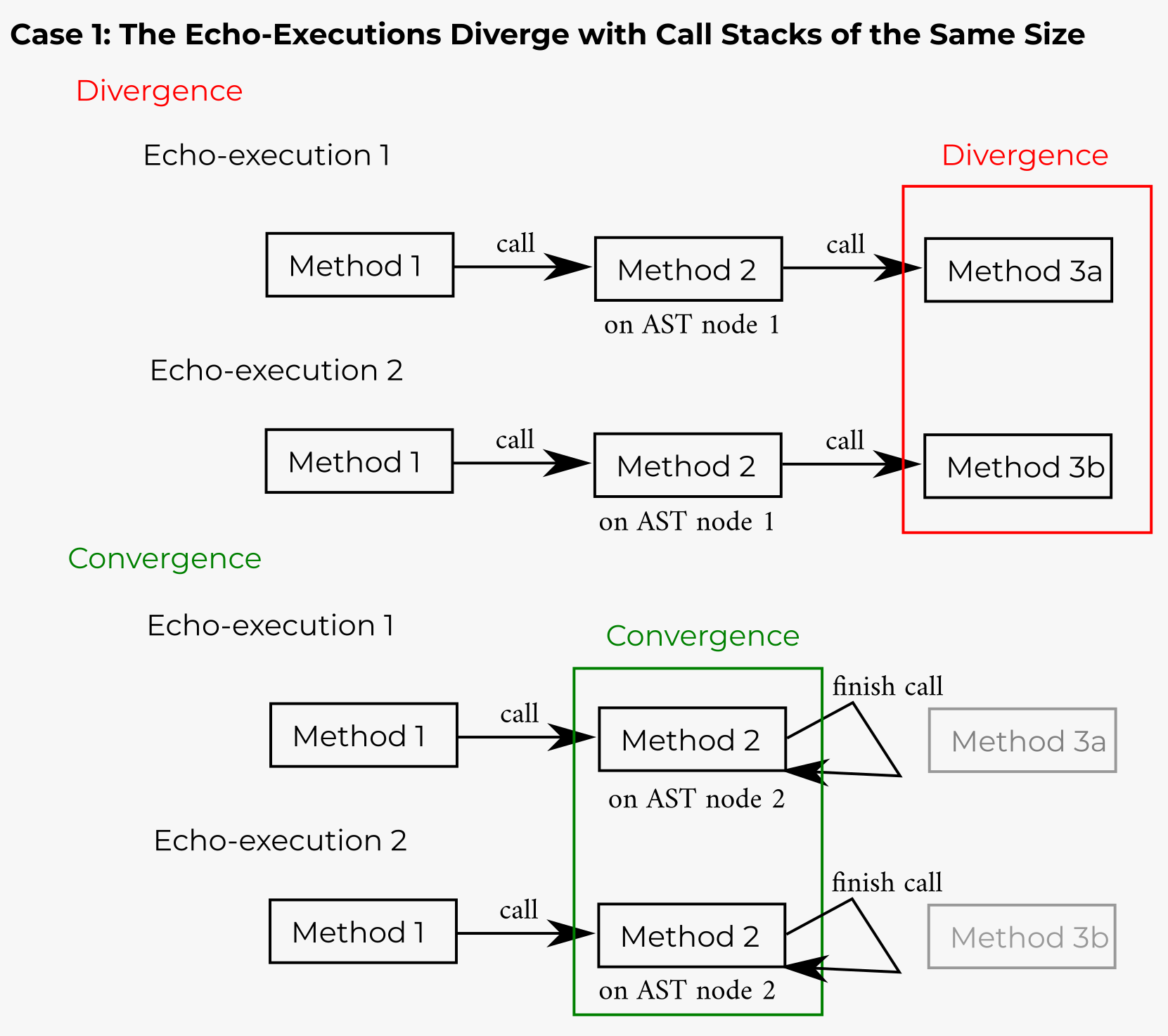}
\caption{When the two echo-executions diverge and have call stacks of the same size, our algorithm steps the echo-executions to finish the current method call and go back to the last method call in which the execution were convergent. It then compares the current AST nodes of the two echo-executions. If it is the same (as is the case in this figure where \ct{AST node 2} = \ct{AST node 2}), the echo-executions have converged. Otherwise, the algorithm repeats the process by finishing the call to \ct{method 2}, comparing the current AST nodes.
\sd{there is too much text, figures should be described by the text and both text and figure should be independently self-sufficient }}
\label{fig:howToReconvergeCase1}
\end{figure}
\begin{figure}[!ht]
\includegraphics[width=\linewidth]{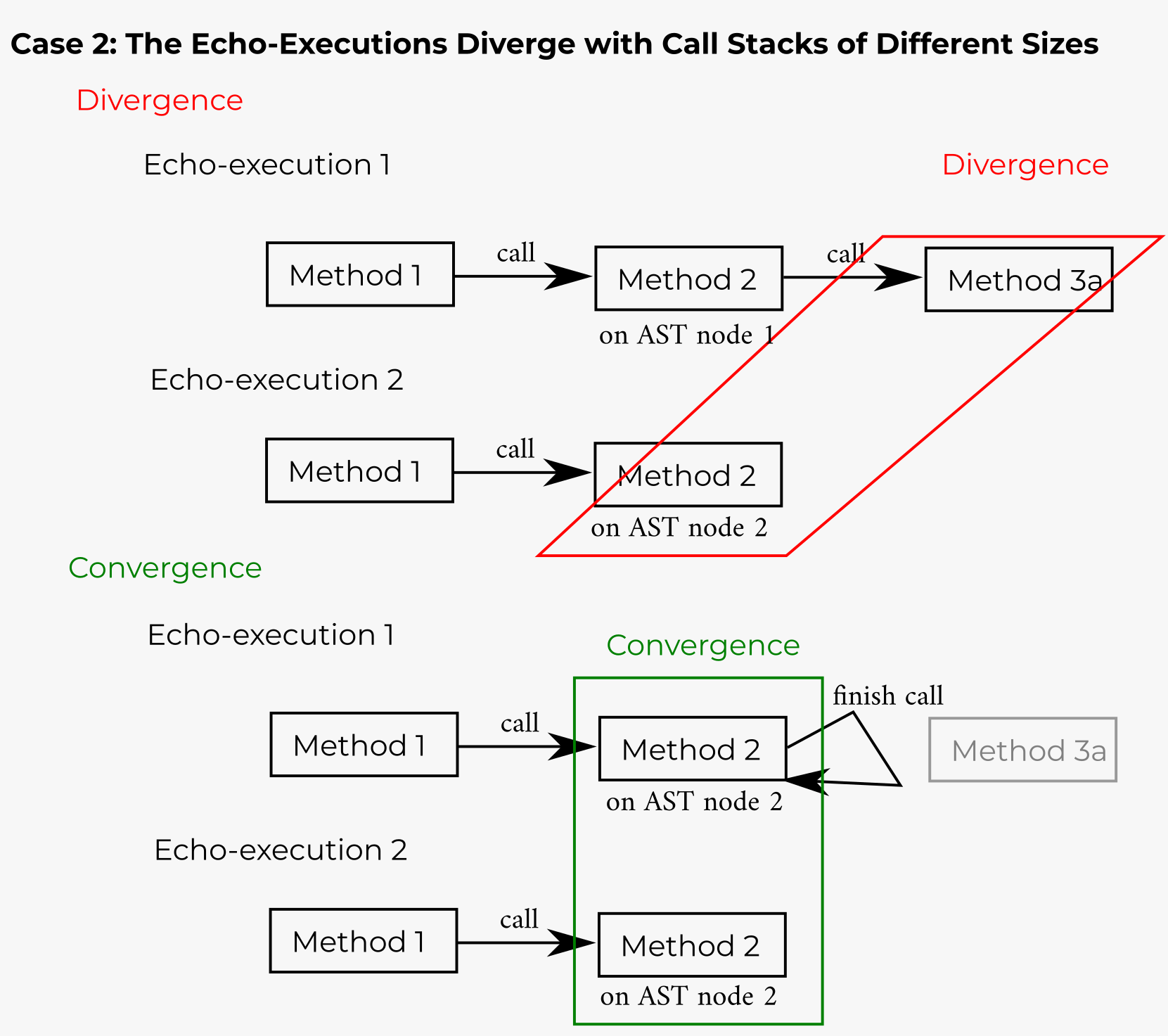}
\caption{When the two echo-executions diverge and have call stacks of different size, our algorithm only finishes the current method call of the echo-execution with the longest call stack.}
\label{fig:howToReconvergeCase2}
\end{figure}

\paragraph{Special case when stepping to the next convergence.}
Sometimes, the echo-executions diverge but their call-stack do not have the same size \sd{I do not get it
if they diverge this is normal that the call-stack do not have the same size}. This can for example happen when the source code change between the two program versions turned a normal method into a Virtual Machine primitive method. When stepping into a primitive method, the VM automatically executes it and returns to the caller. This means that the execution with the normal method is currently one-step-deep into that method, but the other execution is already back in the caller method. To find a convergence in these cases, our algorithm only finishes the current method call of the echo-execution with the longest call stack (See Figure~\ref{fig:howToReconvergeCase2}). For comparison, the normal case is shown in Figure~\ref{fig:howToReconvergeCase1}.

\paragraph{Comparing AST nodes.}
A fundamental part of the CDM algorithm is comparing the AST nodes the echo-executions are executing to determine whether their control-flows have diverged. Since the goal of the CDM algorithm is to find control-flow divergences, it also has to take into account the method and class the AST nodes belong to. For example, two \textit{1+1} AST nodes are equal (in the $=$ sense), but if they are from different methods/classes, we consider them different for the purpose of control-flow. Therefore, the CDM algorithm requires some form of identity ($==$) operator to compare the AST nodes.
However, the standard identity operator cannot be used because the AST nodes to compare are coming from different runtimes. To get them into the controller runtime for the comparison, they would have to go through serialisation and materialisation. These materialised object are always different with regards to identity.
Our solution is to design a new equality operator on remote AST nodes. This operator considers the four properties listed below, and compares them with the equality operator ($=$). Two remote AST nodes sharing these properties means that they come from the same method of the same class, are of the same type and correspond to the same part of the source code. This fits the need of the CDM algorithm for an AST node identity operator checking whether the control-flow of the echo-executions have diverged.
\begin{itemize}
	\item \textbf{methodSelector}: the name of the method this AST node is from.
	\item \textbf{class}: the name of the class the method containing this AST node is from.
	\item \textbf{source}: the source code covered by this AST node. For example, \ct{Point new} for the message node representing the send of the \ct{new} message to the \ct{Point} class.
	\item \textbf{nodeType}: whether this AST node is a message node, a literal node...
\end{itemize}

\begin{figure*}[!ht]
	\includegraphics[width=0.9\linewidth]{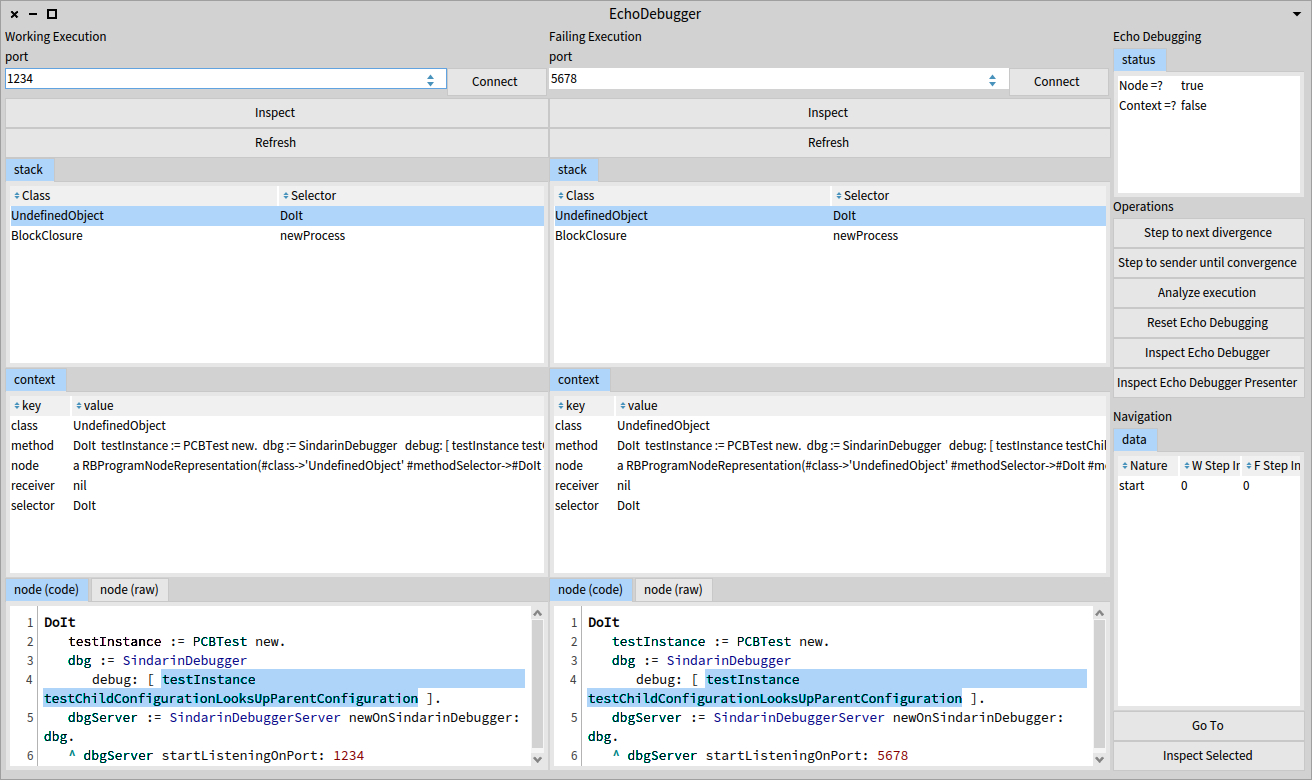}
	\caption{Echo-debugger opened on the Pillar Configuration Bug.}
	\label{fig:echodebuggeronpcbbug1}
\end{figure*}

\section{Example: The Pillar Configuration Bug}
\label{sec:example-pillar-configuration-bug}

Pillar\footnote{\url{https://github.com/pillar-markup/pillar}} is a markup syntax and a tool-suite to generate documentation, books, websites and slides \cite{Duca05h,Arlo16a}. In this section, we use the echo-debugger on a simplified version of a bug encountered in pillar: the pillar configuration bug~\cite{Cost18b}.

\subsection{Starting Knowledge about the Pillar Configuration Bug}

Pillar uses nested configurations to store properties such as the authors, title, default folder and options for the generation given by the users. 
In addition, each file may override new properties (such as authors in a collection of articles).
Each configuration is an environment \ie a dictionary of properties, and has a parent configuration.
Asking a configuration for a given key \textit{key1} is done by sending the message \stCode{key1} to the configuration.
This message is meant not to be understood by the configuration, to call its \stCode{doesNotUnderstand:} method \footnote{This implementation was changed and is not available anymore in recent Pillar distributions because it was a bad idea according to Pillar maintainers.}.
The \stCode{doesNotUnderstand:} method calls the \stCode{lookupProperty:} method of the configuration.
The \stCode{lookupProperty:} methods performs the lookup in the property dictionary of the configuration.
If this dictionary does not contain \textit{key1}, then the \stCode{lookupProperty:} method of the parent configuration is called...

\subsection{The test and the source code change}

The test we are interested in is shown in listing~\ref{lst:testPCB}.
In this test, we create a first configuration \stCode{c1} (line 3) and set the value of its \stCode{mySetting} key to $0$ (line 4).
We then create a second configuration \stCode{c2} (line 5) and declare \stCode{c1} as its parent configuration (line 6).
Finally, we assert that the value of configuration \stCode{c2} for the \stCode{mySetting} key should be $0$, because it should inherit this value from \stCode{c1}.

\begin{lstlisting}[label=lst:testPCB, caption=Test highlighting the Pillar Configuration Bug]
PCBTest>>#testChildConfigurationLooksUpParentConfiguration
| c1 c2 |
c1 := PCBConfig new.
c1 mySetting: 0.
c2 := PCBConfig new.
c2 parentConfig: c1.
self assert: c2 mySetting equals: 0
\end{lstlisting}

This test originally passes, but fails after the following source code change: the developer adds an instance variable to the \stCode{PCBConfig} class, with a getter and a setter method.
Without knowing that the name was already used for a property, the developer names this variable \stCode{mySetting}.
After this change, the test fails, with the message that the property \stCode{mySetting} of c2 is \stCode{nil} instead of $0$.
The test fails because the lookup of \stCode{mySetting} on c2 now returns the value of the \stCode{mySetting} variable (\stCode{nil}) instead of calling the \stCode{doesNotUnderstand:} method as it used to.

\subsection{Echo-debugging the Pillar Configuration Bug}

\paragraph{Setup}
We run three Pharo runtimes in which we loaded the Pillar program~\footnote{\url{https://github.com/dupriezt/PillarConfigBug_Working}} and the echo-debugger with its companion communication packages\footnote{\url{https://github.com/dupriezt/DebuggerCommunication}}. 
We then have:
\begin{description}
\item{\textbf{Working runtime.}} A \emph{working runtime} in which no other code is loaded.
\item{\textbf{Failing runtime.}} A \emph{failing runtime} in which in addition we loaded the breaking changes\footnote{\url{https://github.com/dupriezt/PillarConfigBug_Failing}}.
\item{\textbf{Controller runtime.}} A \emph{controller runtime}. This is from this controller runtime that we will drive the echo-debugging session.
\end{description}

After connecting the runtimes and launching the echo-debugger, as described in the setup process detailed in Section~\ref{sec:setup-echo-debugging-session}, we see the echo-debugger UI shown in Figure~\ref{fig:echodebuggeronpcbbug1}.

\paragraph{Running the CDM algorithm} In the control zone, clicking the \textit{analyze execution} button triggers the CDM algorithm described in Section~\ref{sec:cdm-algorithm}. The result of the CDM algorithm is shown in Figure~\ref{fig:echodebuggeronpcbbug2}.

\begin{figure*}[!ht]
		\includegraphics[width=\linewidth]{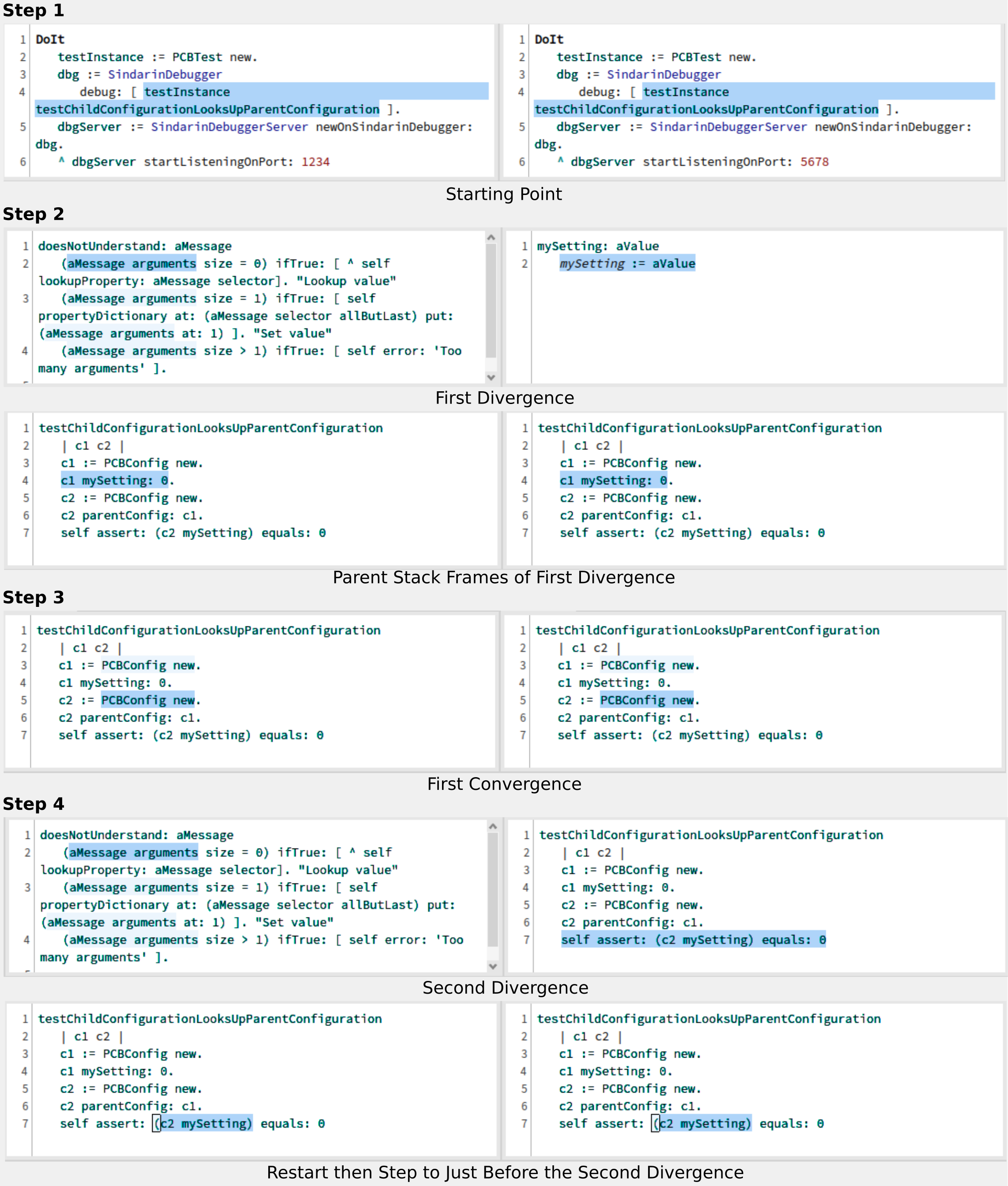}
		\caption{Investigating the echo-executions of the Pillar Configuration Bug}
		\label{fig:echodebuggeronpcbbugstepscompilation}
	\end{figure*}

\paragraph{Investigating the echo-executions}
Now we explain step by step how the echo-executions help us find the root cause of the problem. Figure~\ref{fig:echodebuggeronpcbbugstepscompilation} contains the relevant screenshots for the steps listed below, marked in bold in the text.

\begin{enumerate}
	\item  \textbf{Starting Point.} This is the setup code that has been executed to instantiate a Sindarin debugger on the test execution. The highlighted statement, about to be executed, is the test execution itself. In this figure and all the similar ones, the working execution is shown on the left, while the failing execution is shown on the right.

	\item We step both echo-executions to the \textbf{First Divergence}. The W execution is in a \stCode{doesNotUnderstand:} method, while the F execution is in the \stCode{mySetting:} setter method that was added by the source code change. On the \textbf{Parent of these Stack Frames}, we see that the test executions are setting the value of the \stCode{mySetting} property in configuration \stCode{c1}. We deduce that the configuration did not understand the \stCode{mySetting:} message in the W execution, but it did in the F execution.  The developer already expects this, since he just added the \stCode{mySetting:} setter method on purpose.
	
%
	
	\item We step both echo-executions to the \textbf{First Convergence}. We see that after the execution of the \stCode{mySetting:} message was different between the echo-executions, they reconverge at the next statement of the test method. Notice that the W execution took 106 steps to reach this convergence, while the F execution only took 9.

%

	\item We step both echo-executions to the \textbf{Second Divergence}.
	Here, the F execution is about to execute the whole assertion of the test, while the W execution is in a \stCode{doesNotUnderstand} method.
	To have a better look, we can \textbf{Restart} both echo-executions and step them until they reach the step just before this divergence event (114 steps for the W execution, 17 for the F execution).
	We see that both executions were about to execute the \stCode{c2 mySetting} statement of the test assertion.
	We deduce that this call resulted in a \stCode{doesNotUnderstand:} in the W execution, while it resulted in the \stCode{mySetting} getter method being called in the F execution.
	Using the debuggers, we separately inspect the two echo-executions from this point. In the W execution, doing a few steps shows the configuration \stCode{c2} not understanding the message \stCode{mySetting}, looking up its property dictionary, and delegating the lookup to its parent configuration.
	In the F execution, we inspect the \stCode{c2} configuration object to find that the value of its \stCode{mySetting} instance variable is \stCode{nil}.


	\item We found the cause of the bug: adding a getter for \stCode{mySetting} on the pillar configuration class caused it to understand the \stCode{mySetting} message. This prevented the property lookup from escalading to the parent configuration. \sd{will the reader really understand here?}
\end{enumerate}

\section{Discussion}
\label{sec:discussion}



\paragraph{State differences.}
A limitation of our solution is that it only considers differences between the echo-executions in terms of control-flow. While such differences are important and helpful, differences in terms of \textit{state} may also be very helpful to the developer. For example, recognizing when the echo-executions have the same control-flow but act on objects with different states. \sd{How would that help (it will but how)?}

\paragraph{Back-in-time debugging.}
After the CDM algorithm presented in Section~\ref{sec:cdm-algorithm} has fully run both executions to detect when divergence and convergence events occur, the echo-debugger restarts the echo-executions and steps them forward to reach the events the developer wants to inspect. This is a rudimentary form of back-in-time debugging, which assumes that the echo-executions are deterministic. More advanced techniques of back-in-time debugging~\cite{Lewi03b, Hofe06b, Poth07a, Lien08b} could be used to remove this assumption.

\paragraph{Optimization of the CDM algorithm.}
In this paragraph, we discuss an implementation detail that proved critical in terms of performance.
While our initial implementation of the CDM algorithm was almost instantaneous for small executions (around 250 steps), it was very slow for larger executions (more than 1 hour for around 5 million steps). The biggest performance bottleneck were the HTTP requests between the controller and echo-runtimes.
The naïve implementation of the CDM algorithm sends many small HTTP requests to the echo-runtimes. Among others, one request per step, one request per AST node comparison to get the AST node, and one request each time the size of the call stack is needed.\\
To reduce the number of HTTP requests necessary, we simplified the data needed by the CDM algorithm running in the controller runtime. With this simplification, we no longer need the echo-executions to run in parallel. Instead, the echo-runtimes fully run their echo-executions locally, collecting the necessary data, and then send this data in big batches to the controller runtime. The controller runtime then performs the CDM algorithm offline on the data.\\
Data simplification: since the CDM algorithm only compares AST nodes to each other, it does not need the full dictionary representation of these nodes, and can work simply with the hashes of these representations. Also, the CDM algorithm does not need the complete call-stacks of the echo-executions, it only needs their size. With these two simplifications, the echo-runtimes fully run their echo-execution locally with no intervention from the controller runtime. After each execution step, they log a) the hash of the dictionary representation of the current AST node and b) the size of the call-stack.\\
These optimizations reduced the time necessary to run the CDM algorithm on an execution around 5 million steps long from more than an hour to 2 minutes.

\section{Related Works}
\label{sec:related}

\paragraph{Test inputs.}
Palikareva \etal~\cite{Pali16a} describe a technique called \textit{Shadow symbolic execution}, designed to generate test inputs that cover new program behavior introduced by a patch. This technique symbolically executes a test in both program versions (before and after the patch) and compares these executions to find test inputs that lead to new behavior in the patched program and should be tested. This technique requires the developer to manually annotate the program to merge the old and new versions of the code. The Echo-Debugger does not have this requirement.

Brumley \etal~\cite{Brum07a} solve logical formulae created from two different implementations of the same protocol (for example HTTP) to find deviations: inputs such that the output of the two implementations are semantically different. This technique produces inputs generating deviations between two implementations. By contrast, the Echo-Debugger is a tool to explore how two programs deviate on a given execution.

\paragraph{Delta Debugging and compared execution.}
Zeller~\cite{Zell99a, Zell02b} presents the \textit{Delta debugging} algorithm. This algorithm takes 2 versions of a program, and a test that was passing in the old version, but is failing in the new version. Delta debugging uses a divide-and-conquer approach to try multiple subsets of the code change and find the smallest subset that turns the test from green to red.

Abramson \etal~\cite{Abra95a} propose \textit{relative debugging}, a paradigm where the developer formulates a set of equality assertions about key data structures in the old and new versions of a program.  The relative debugger is responsible for executing the two program versions in parallel and report any difference between the marked data structures.  The two major differences with the Echo-Debugger are that 1) relative debugging deals with state differences while the Echo-Debugger deals with control-flow differences and 2) relative debugging requires manual interventions of the developer to mark the data structures they want to compare, and at which lines of code in the two programs to perform the comparison.

In the WhyLine~\cite{Ko04a,Ko08a} tool, the developer asks questions about a recorded execution. 
The tool exploit traces to answer the questions, and tell why a particular variable has or has not a given value. Recorded divergences and convergences in the echo-debugger could be leveraged to ask questions about the execution in order to bring a better understanding of why two execution diverge.

Pinocchio~\cite{Verw10a} is a proof-of-concept implementation of a first-class code interpreter. Developers subclass the default interpreter to add behaviors to the code execution. An example use case is the creation of a \textit{parallel debugger}, running two interpreters in parallel and comparing their state after each step. As opposed to the Echo-Debugger, the two interpreters of Pinocchio runs in the same runtime, and can only compare two executions on the same code base.

\paragraph{Algorithmic debugging.}
Algorithmic debugging is a technique proposed in 1982 by E. Y. Shapiro in the context of logic programming~\cite{Shap82a, Silv11a}.
Algorithmic debugging requires an oracle to compare execution outputs.
These techniques try to isolate faulty code based on how developers assert the outputs of faulty and successful executions.
An oracle could be used in the echo-debugger, to ask the developer to assert if a given convergence or divergence is normal or not (\textit{e.g.}, between two program versions).
This would help to focus on convergences and divergences that are relevant for the user, and ignore mundane differences like semantic-preserving refactoring.

\section{Future Work}
\label{sec:future}

As future work to expand the echo-debugger presented in this paper, we identified 3 main axis.

\paragraph{State Differences.} The Echo-debugger presented in this paper focuses on control-flow differences between the echo-executions. Another dimension in which two executions can differ is state (for example in the content of their variables). Incorporating state differences into the Echo-debugger, possibly inspired by the work of Henry Liberman~\cite{Lieb84a}, will make it  paint a more complete picture of the differences between the executions.

\paragraph{Automated Setup.} Setting up an echo-debugging session is a multi step process that can be tedious. An improvement axis consists in developing an automated setup tool to create the three runtimes, load the echo-debugger and its dependencies, run the debugger servers and client, and link them over HTTP. This tool could for example take as input a link to a git repository and two commit ids.

\paragraph{Using a Back-in-time Debugger as Back-end.} Back-in-time debuggers are specifically designed to allow faithful replays of executions. The Echo-debugger requires this feature, and currently implements it by naïvely replaying the executions. This works for deterministic, isolated executions, but not for more complex executions. Using a back-in-time debugger as back-end will lift this limitation of the Echo-Debugger. 

\section{Conclusion}
\label{sec:conclusion}


In this paper, we tackled the challenge of debugging two similar executions in parallel.
We proposed the echo-debugger: an interactive debugger to debug two similar executions running in different runtimes.
We also proposed the Convergence Divergence Mapping algorithm (CDM), an algorithm that fully runs both executions and compares the AST nodes they are executing to build a map of when they diverge and converge in terms of control-flow.
This map records how many steps each echo-execution took to reach each event.
The echo-debugger can then restart the echo-executions and step them to any event of this map the developer wants to inspect.
We showed on an example how the echo-debugger helps finding the cause of a vicious bug.

The main limitation of the echo-debugger is that it focuses on the control-flow differences between the executions, but ignores the potential difference of state. This constitutes the main improvement direction of the echo-debugger.
Additionally, since the echo-debugger should be able to replay an execution, the execution should be 
deterministic. Combining it with a back-in-time debugger would lift this limitation.

%


\subsection*{Acknowledgements} 
This work was supported by Ministry of Higher Education and Research, Nord-Pas de Calais Regional Council, CPER Nord-Pas de Calais/FEDER DATA Advanced data science and technologies 2015-2020.

%

\bibliographystyle{abbrv}
\bibliography{rmod,others}
\end{document}

%% file: macros.tex
\usepackage{xcolor}
\usepackage[normalem]{ulem} 			
\usepackage{ifthen}

\newboolean{showcomments}
\setboolean{showcomments}{false}

\ifthenelse{\boolean{showcomments}}
{
	\newcommand{\nb}[3]{
		{\colorbox{#2}{\bfseries\sffamily\scriptsize\textcolor{white}{#1}}}
		{\textcolor{#2}{\textsf\small$\blacktriangleright$\textit{#3}$\blacktriangleleft$}}}
	 
	\newcommand{\bnote}[2]{\fbox{\color{blue}\bfseries\sffamily\scriptsize#1}
    	{\color{blue}\textsf\small$\blacktriangleright$\textit{#2}$\blacktriangleleft$}}
	\newcommand{\old}[1]{{\color{gray}\sout{#1}}} 
	\newcommand{\del}[1]{\old{#1}} 
	\newcommand{\ins}[1]{{\textcolor{blue}{\uline{#1}}}} 
	\newcommand{\ugh}[1]{{\textcolor{red}{\uwave{#1}}}} 
	\newcommand{\chg}[2]{{\textcolor{red}{\sout{#1}}}{\ra}\textcolor{blue}{\uline{#2}}} 
	
	\newcommand{\fix}[1]{\bnote{FIX}{#1}}
}{
	\newcommand{\bnote}[2]{}
	\newcommand{\nb}[3]{}
	\newcommand{\old}[1]{}
	\newcommand{\del}[1]{}
	\newcommand{\ins}[1]{}
	\newcommand{\ugh}[1]{}
	\newcommand{\chg}[2]{}
	
	\newcommand{\fix}[1]{}
}

\newcommand{\hide}[1]{}

\newcommand{\td}[1]{\nb{Thomas}{blue}{#1}}

\newcommand{\sd}[1]{\nb{Stephane}{brown}{#1}}

\graphicspath{{figures/}}

\newcommand{\commented}[1]{}

\newcommand{\stCode}[1]{\textsf{#1}}

\newcommand{\ie}{\emph{i.e.,}\xspace}
\newcommand{\etal}{\emph{et al.,}\xspace}
\newcommand{\ct}[1]{{\textsf{#1}}\xspace}

\usepackage{url}
\makeatletter
\def\url@leostyle{%
  \@ifundefined{selectfont}{\def\UrlFont{\textsf}}{\def\UrlFont{\small\sffamily}}}
\makeatother
\usepackage{etoolbox}

\newtoggle{InString}{}
\togglefalse{InString}
\usepackage{color}
\usepackage{textcomp}
\usepackage{listings}

\definecolor{source}{gray}{0.85}
\newcommand{\myCommentStyle}[1]{{\color{gray}\itshape #1}}
\newcommand{\myStringStyle}[1]{{#1}}
\newcommand{\mySymbolStyle}[1]{{#1}}
\newcommand{\myKeywordStyle}[1]{{#1}}
\newcommand{\myGlobalStyle}[1]{{\bfseries #1}}

\makeatletter

\usepackage{ifluatex}
\ifluatex
  \usepackage{pdftexcmds}
  \let\pdfstrcmp\pdf@strcmp
\fi

\newcommand*\idstyle[1]{%
  \expandafter\id@style\the\lst@token{#1}\relax%
}
\def\id@style#1#2\relax{%
  \ifnum\pdfstrcmp{#1}{\#}=0%
    \mySymbolStyle{\the\lst@token}%
  \else%
    \edef\tempa{\uccode`#1}%
    \edef\tempb{`#1}%
    \ifnum\tempa=\tempb%
      \myGlobalStyle{\the\lst@token}%
    \else%
      \the\lst@token%
    \fi%
  \fi%
}
\makeatother

\lstnewenvironment{code}{%
  \noindent%
  \minipage{\linewidth}%
}{%
  \endminipage%
}%
\lstnewenvironment{codeWithLineNumbers}{%
  \lstset{numbers=left}%
  \noindent%
  \minipage{\linewidth}%
}{%
  \endminipage%
}%


%% file: EchoDebugging_IWST_HalUpload.bbl
\begin{thebibliography}{10}

\bibitem{Abra95a}
D.~Abramson, I.~Foster, J.~Michalakes, and R.~Sosic.
\newblock Relative debugging and its application to the development of large
  numerical models.
\newblock In {\em In proceedings of IEEE supercomputing}, 1995.

\bibitem{Arlo16a}
T.~Arloing, Y.~Dubois, D.~Cassou, and S.~Ducasse.
\newblock Pillar: A versatile and extensible lightweight markup language.
\newblock In {\em International Workshop on Smalltalk Technologies {IWST'16}},
  Prague, Czech Republic, Aug. 2016.

\bibitem{Blac09a}
A.~P. Black, S.~Ducasse, O.~Nierstrasz, D.~Pollet, D.~Cassou, and M.~Denker.
\newblock {\em Pharo by Example}.
\newblock Square Bracket Associates, Kehrsatz, Switzerland, 2009.

\bibitem{Brum07a}
D.~Brumley, J.~Caballero, Z.~Liang, J.~Newsome, and D.~Song.
\newblock Towards automatic discovery of deviations in binary implementations
  with applications to error detection and fingerprint generation.
\newblock In {\em In Proceedings of the USENIX Security Conference. USENIX
  Association}, 2007.

\bibitem{Cost18b}
S.~Costiou.
\newblock {\em {Unanticipated behavior adaptation : application to the
  debugging of running programs}}.
\newblock Theses, {Universit{\'e} de Bretagne occidentale - Brest}, Nov. 2018.

\bibitem{Duca05h}
S.~Ducasse, L.~Renggli, and R.~Wuyts.
\newblock {SmallWiki} --- a meta-described collaborative content management
  system.
\newblock In {\em Proceedings ACM International Symposium on Wikis
  (WikiSym'05)}, pages 75--82, New York, NY, USA, 2005. ACM Computer Society.

\bibitem{Dupr19a}
T.~Dupriez, G.~Polito, S.~Costiou, V.~Aranega, and S.~Ducasse.
\newblock Sindarin: A versatile scripting api for the pharo debugger.
\newblock In {\em DLS'19, Dynamic Language Symposium}, 2019.

\bibitem{Hofe06b}
C.~Hofer.
\newblock Implementing a backward-in-time debugger.
\newblock Master's thesis, University of Bern, Sept. 2006.

\bibitem{Ko04a}
A.~J. Ko and B.~A. Myers.
\newblock Designing the whyline: a debugging interface for asking questions
  about program behavior.
\newblock In {\em Proceedings of the 2004 conference on Human factors in
  computing systems}, pages 151--158. ACM Press, 2004.

\bibitem{Ko08a}
A.~J. Ko and B.~A. Myers.
\newblock Debugging reinvented: Asking and answering why and why not questions
  about program behavior.
\newblock In {\em In Proceedings of the 30th International Conference on
  Software Engineering, ICSE 08}, 2008.

\bibitem{Lewi03b}
B.~Lewis.
\newblock Debugging backwards in time.
\newblock In {\em Proceedings of the Fifth International Workshop on Automated
  Debugging (AADEBUG'03)}, Oct. 2003.

\bibitem{Lieb84a}
H.~Lieberman.
\newblock Steps towards better debugging tools for lisp.
\newblock {\em ACM Symposium on Lisp and Functional Programming}, 1984.

\bibitem{Lien08b}
A.~Lienhard, T.~G\^irba, and O.~Nierstrasz.
\newblock Practical object-oriented back-in-time debugging.
\newblock In {\em Proceedings of the 22nd European Conference on
  Object-Oriented Programming (ECOOP'08)}, volume 5142 of {\em LNCS}, pages
  592--615. Springer, 2008.
\newblock {ECOOP} distinguished paper award.

\bibitem{Pali16a}
H.~Palikareva, T.~Kuchta, and C.~Cadar.
\newblock Shadow of a doubt: Testing for divergences between software versions.
\newblock In {\em International Conference on Software Engineering (ICSE
  2016)}, 2016.

\bibitem{Pers17a}
M.~Perscheid, B.~Siegmund, M.~Taeumel, and R.~Hirschfeld.
\newblock Studying the advancement in debugging practice of professional
  software developers.
\newblock {\em Software Quality Journal}, 25(1):83--110, 2017.

\bibitem{Poth07a}
G.~Pothier, E.~Tanter, and J.~Piquer.
\newblock Scalable omniscient debugging.
\newblock {\em Proceedings of the 22nd Annual SCM SIGPLAN Conference on
  Object-Oriented Programming Systems, Languages and Applications (OOPSLA'07)},
  42(10):535--552, 2007.

\bibitem{Shap82a}
E.~Y. Shapiro.
\newblock Algorithmic program diagnosis.
\newblock In {\em Proceedings of Principles of Programming Languages (Popl)},
  pages 299--308. ACM, 1982.

\bibitem{Silv11a}
J.~Silva.
\newblock A survey on algorithmic debugging strategies.
\newblock {\em Advances in engineering software}, 42(11):976--991, 2011.

\bibitem{Somm01}
I.~Sommerville.
\newblock {\em Software Engineering (6th ed.)}.
\newblock Addison-Wesley, 2001.

\bibitem{Verw10a}
T.~Verwaest, C.~Bruni, D.~Gurtner, A.~Lienhard, and O.~Niestrasz.
\newblock Pinocchio: bringing reflection to life with first-class interpreters.
\newblock In {\em OOPSLA '10: Proceedings of the ACM international conference
  on Object oriented programming systems languages and applications}, 2010.

\bibitem{Wong16a}
E.~W. Wong, R.~Gao, R.~Abreu, and F.~Wotawa.
\newblock A survey on software fault localization.
\newblock {\em IEEE Transactions on Software Engineering}, 42(8):707--740,
  2016.

\bibitem{Zell99a}
A.~Zeller.
\newblock Yesterday, my program worked. today, it does not. why?
\newblock In {\em ESEC/FSE-7: Proceedings of the 7th European software
  engineering conference held jointly with the 7th ACM SIGSOFT international
  symposium on Foundations of software engineering}, pages 253--267, London,
  UK, 1999. Springer-Verlag.

\bibitem{Zell02b}
A.~Zeller.
\newblock Isolating cause-effect chains from computer programs.
\newblock In {\em SIGSOFT '02/FSE-10: Proceedings of the 10th ACM SIGSOFT
  symposium on Foundations of software engineering}, pages 1--10, New York, NY,
  USA, 2002. ACM Press.

\bibitem{Zell05a}
A.~Zeller.
\newblock {\em Why Programs Fail: A Guide to Systematic Debugging}.
\newblock Morgan Kaufmann, Oct. 2005.

\end{thebibliography}
